\documentclass[12pt]{article}
\usepackage{epsfig,amsfonts,amssymb}
\usepackage{hyperref}

\usepackage{cite}
\usepackage{bm}
\usepackage[table]{xcolor}

\usepackage{amsmath}
\usepackage{bigints}

\usepackage[mathscr]{euscript}
 \let\mathscr\relax

\begin{document}
\baselineskip 24pt

\begin{center}

{\Large \bf Supercritical linear dilaton capping state using Interpolating functions}

\end{center}

\vskip .6cm
\medskip

\vspace*{4.0ex}

\baselineskip=18pt

\centerline{\large \rm Ritabrata Bhattacharya$^*$}

\vspace*{4.0ex}

\centerline{\large $^*$Chennai Mathematical Institute,}
\centerline{\large H1, SIPCOT IT Park, Siruseri, Kelambakkam 603103, India}

\begin{NoHyper}
  \let\thefootnote\relax\footnotetext{electronic address: {\tt {ritabratab@cmi.ac.in} }}
\end{NoHyper}

\vspace*{5.0ex}

\thispagestyle{empty}
\centerline{\bf Abstract} \bigskip
We consider time dependent backgrounds with a time-like linear dilaton which leads to a weakly coupled theory at assymptotic future known as the Supercritical Linear Dilaton (SCLD) phase. Even after projecting out the tachyon there are naive divergences in the partition function due to the spectrum of relevant deformations. Although the partition function seems divergent, the background is actually well defined in the far future due to vanishing backreactions, provided we start from an appropriate initial state called the ``capping" state. We use the technique of interpolating functions to construct this state by embedding the SCLD phase in a strong coupling completion of string theory. Although we are unable construct a specific example, we do provide the general strategy in cases where S-duality may provide the completion.

\vfill \eject

\baselineskip 18pt

\setcounter{page}{1}
\tableofcontents
\section{Introduction}
Our universe as we know is an expanding one. So to describe a consistent quantum theory of gravity at early times i.e. near the big bang singularity one requires the solutions describing quantum gravity in time dependent backgrounds. String theory in Linear Dilaton backgrounds provide a way to tackle this problem in a somewhat systematic manner. In \cite{Antoniadis:1988aa},\cite{Polchinski:1989fn},\cite{Antoniadis:1988vi},\cite{Antoniadis:1990uu} it was noticed that time-like linear dilaton solutions of supercritical string theories are strongly coupled at assymptotically early times and hence difficult to solve. Although at assymptotically late times one has much more control since there, the theories become weakly coupled and one can solve them perturbatively. In this regard we should mention that we take the following conventions for the dilaton $\Phi$ and the string coupling $g_{st}$,
$$\Phi=-QX^0\ ,\quad g_{st}=e^{\langle\Phi\rangle}=e^{-QX^0}\ ,\ \ \text{with}\ \ Q>0\ .$$
In \cite{Aharony:2006ra}, the authors found that in such cases the spectrum of relevant deformations of the ``matter CFT" has modes which grow in time although their backreaction on the geometry is small at assymptotically late times due to the small string coupling. The condition for the backreaction being small is stated below but one should note that these are not the usual tachyons we are familiar with. These modes were termed ``\textit{Pseudo-tachyons}" in the literature. Tachyonic solutions also grow in time but they have large back reactions on the geometry causing instabilities in the background. If we consider the deformations,
$$\int d^2z\ \sum_i\mu_i\mathcal{O}_ie^{\kappa_iX^0}$$
with $\mathcal{O}_i$ having dimension $\Delta_i$. This is marginal if,
$$\Delta_i+\frac{1}{4}\kappa_i(\kappa_i+2Q)=2\ \Rightarrow\ \kappa_i=-Q\pm\sqrt{Q^2-4(\Delta_i-2)}\ .$$
For $2<\Delta_i<2+Q^2/4$ we see that the properly normalised relevant mode goes like,
$$\tilde{\lambda}=\frac{\lambda}{g_{st}}\sim e^{\pm QX^0\sqrt{1-4(\Delta_i-2)/Q^2}}\ .$$
So we see that one of the modes grow in time and these are the ones identified as \textit{Pseudo-tachyons}.

The condition that the pseudo-tachyons have small backreaction and hence provide a well defined i.e. stable background, is that we begin with an appropriate initial state. This is sometimes called a ``capping" state. As per current understanding, unambiguous description of such states is still an open problem. Although some work has been done in this regard, where smooth evolution from ``nothing" was considered \cite{McGreevy:2005ci}. Basically the authors studied the effect of closed string tachyon condensation on the spacetime with a space-like singularity at the level of general relativity. 

Another way to identify the ``capping" state is to embed the Supercritical Linear Dilaton(SCLD) phase in a strong coupling completion of string theory. The goal of the current paper is to take this route for identifying the capping state. First let us look at the obstacles we face in achieving this goal, then we mention cases where there is a systematic way to circumvent these problems to obtain a reasonably accurate result.

\subsubsection*{Obstacles}
\begin{enumerate}
\item{The most obvious obstactle is that we do not know whether there exists a consistent strong coupling completion for supercritical string theories.}
\item{Even if we are able find certain examples where such completion is possible, computing physical quantites relevant for the target spacetime is considerably difficult since spacetime supersymmetry is broken due to the linear dilaton which couples to the world-sheet curvature.\\
For example suppose we want to compute the renormalised mass of different operators at a given mass level. In presence of supersymmetry it is sufficient to compute the renormalised mass of one component of the supermultiplet or superfield, the rest are fixed by supersymmetry. This will obviously not be true if supersymmetry is broken.}
\item{Full non-perturbative treatment of string theory is yet to be as well understood as perturbative string theory which is why even if we resolve the first obstacle, the resolution may be true only in some special cases.}
\end{enumerate}
We argue that if we have two supercritical theories which are S-dual to each other describing the weak and strong coupling regime then we can attempt to construct the capping state within reasonable accuracy using the technique of interpolating functions. Although we fail to find a specific example and produce numbers, it may help to know the strategy of constructing the capping state in case we suceed in finding an example. 

The technique of interpolating function involves computing the renormalized mass function $F(g)$ as a perturbative expansion in string coupling. If we have two theories at the weak and strong coupling end which are S-dual to each other, then we have two functions $F_W$ and $F_S$ which have perturbative expansions in the weak and strong coupling respectively at the two ends. We will closely follow the approach described in \cite{Sen:2013oza} to find an interpolating function $F_{m,n}(g)$ which matches $F_W$ upto $m$-th power in the weak coupling expansion and $F_S$ upto the $n$-th power in the strong coupling expansion. This interpolating function is supposed to give the renormalized mass of the theory for the whole range of coupling to a reasonable accuracy. Other approaches are also available in the literatre \cite{Kleinert:2001ax} and the Pad\'e approximant approach.\\

The rest of the paper is organised as follows. In Section \ref{sec:LD} we discuss briefly the known results and the conventions we use for linear dilaton backgrounds and mass renormalization in string perturbation theory. We will also derive the partition function for SCLD backgrounds and show that modular invarince dictates that only certain spacetime dimensions are allowed, which is well known. In Section \ref{sec:Q1} we compute the two point amplitude of a pseudo-tachyonic mode on the torus, which gives the renormalized mass at one loop for this mode. In Section \ref{sec:IF} we describe strategy for determining the ``capping" state with the help of interpolating functions. And finally we conclude with some discussions in Section \ref{disc}.

\section{Known results and conventions to be used}
\label{sec:LD}
\subsection{Linear dilaton backgrounds}
In this section we give a brief review of known results in the case when a dilaton which is linear in the string fields, couples to the world-sheet curvature \cite{Polchinski:101b}. So,
\begin{equation}
\phi=V_{\mu}X^{\mu}\ ,
\end{equation} 
and the world-sheet Eucledian action for the strings is given by,
\begin{equation}
S_P=\frac{1}{4\pi\alpha^{\prime}}\int d^2x\sqrt{\gamma}\gamma^{ab}\partial_aX^{\mu}\partial_bX_{\mu}\ +\ \frac{1}{4\pi}\int d^2x\sqrt{\gamma}\phi R
\end{equation}
Varrying this action w.r.t the world-sheet metric $\gamma$ we obtain the energy momentum tensor for this theory. We give the result here for the holomorphic and anti-holomorphic part of the E.M. tensor,
\begin{eqnarray}
T(z)&=& -\frac{1}{\alpha^{\prime}}:\partial X^{\mu}\partial X_{\mu}:\ +\ V_{\mu}\partial^2X^{\mu} \nonumber\\
\bar{T}(\bar{z})&=& -\frac{1}{\alpha^{\prime}}:\bar{\partial} X^{\mu}\bar{\partial} X_{\mu}:\ +\ V_{\mu}\bar{\partial}^2X^{\mu}\ .\label{eq:emLD}
\end{eqnarray}
With this energy momentum tensor we want to find the modified central charge as well as the modifed conformal weights of the vertex operators. Now we want to work with time-like dilaton so we choose $V_{\mu}\equiv(Q,\vec{0})$   and hence we end up with (space-time metric is the minkowski metric)
\begin{equation}
T(z)= -\frac{1}{\alpha^{\prime}}:\partial X^{\mu}\partial X_{\mu}:\ -\ Q\partial^2X^{0}
\end{equation}
and similarly for $\bar{T}(\bar{z})$.

We want to compute the the $T(z)T(0)$ as well as the $T(z)e^{ik.X}$ OPE's. We know that now $\partial X^0$ is no longer a primary while $e^{-ik^0X^0}$ is still a primary. Looking only at the leading order divergences, we get the following results\footnote{Here $D$ is the dimensionality of the target space-time.},
\begin{eqnarray}
T(z)T(0)&\sim & \frac{D-6\alpha^{\prime}Q^2}{2z^4}\ +\ ....\\
T(z)e^{ik.X}(0) &\sim & \left(\frac{\alpha^{\prime}k^2}{4}-\frac{iQk^0\alpha^{\prime}}{2}\right)\frac{e^{ik.X}}{z^2}\ +\ ....
\end{eqnarray}
Thus we have $c=\bar{c}=D-6\alpha^{\prime}Q^2$. And the conformal dimension of the vertex operator, $$h_v=\frac{\alpha^{\prime}k^2}{4}-\frac{iQk^0\alpha^{\prime}}{2}=\frac{\alpha^{\prime}}{4}\left[\vec{k}^2-(k^0)^2-2ik^0Q)\right]$$

\subsection{The one loop partition function}
Before computing the partition function for such theories we include world-sheet supersymmetry so that in the case of $Q = 0$ we have a GSO projected spectrum of critical superstrings in 10 dimension where the tachyon is projected out.

We use the known results for fermionic traces \cite{Polchinski:102b},
\begin{eqnarray}
Z^{\alpha}_{\ \beta}=\text{Tr}_{\alpha}\left[q^{H}\text{exp}(i\pi\beta F)\right]=\frac{1}{\eta(\tau)}\vartheta\begin{bmatrix}
\alpha/2\\
\beta/2
\end{bmatrix}(0|\tau)\ .
\end{eqnarray}
Here $F$ is the world-sheet fermion number, $\alpha$ and $\beta$ take integer values modulo 2. For $\alpha$ even we have trace over the NS and for odd we have trace over R states. 

The total central charge due to the \textit{bc} and $\beta\gamma$ ghosts is given by, $c_g=-26+11=-15$ and the matter CFT has central charge $c_m=D-6\alpha^{\prime}Q^2+\frac{D}{2}$. So for the central charge for the matter + ghost CFT to vanish we need,
\begin{equation}
\frac{3D}{2}-6\alpha^{\prime}Q^2=15\ \ \Rightarrow\ D-4\alpha^{\prime}Q^2=10\ .\label{eq:D_Q}
\end{equation}
For specificity we assume $D/2$ to be odd and we will GSO project onto even world-sheet fermion number to remove the tachyon.

Now we have the following well results on the torus,
\begin{equation}
\langle c(w_1)b(w_2)\bar{c}(\bar{w}_3)\bar{b}(\bar{w}_4)\rangle_{T^2}=|\eta(\tau)|^4\ .
\end{equation}
The scalar partition function ($q=e^{2\pi i\tau}\ \Rightarrow\ q\bar{q}=e^{-4\pi\tau_2}$),
\begin{eqnarray}
Z_s(\tau)&=&(q\bar{q})^{-\frac{D-6\alpha^{\prime}Q^2}{24}}\text{Tr}\left(q^{L_0}\bar{q}^{\bar{L}_0}\right) \\\nonumber\\
Z_s(\tau)&=&V_D(q\bar{q})^{-\frac{D-6\alpha^{\prime}Q^2}{24}}\int\frac{d^{D-1}k}{(2\pi)^{D-1}}\frac{dk^0}{2\pi}\text{exp}\left[-\pi\alpha^{\prime}\tau_2\big(\vec{k}^2-(k^0)^2-2ik^0Q\big)\right]\nonumber\\
&&\hspace*{6cm}\times\prod_{\mu,n}\sum_{N_{\mu n},\bar{N}_{\mu n}}q^{nN_{\mu n}}\bar{q}^{n\bar{N}_{\mu n}}\nonumber\\
Z_s(\tau)&=&iV_D(q\bar{q})^{-\frac{D}{24}}(q\bar{q})^{\frac{\alpha^{\prime}Q^2}{4}}\int\frac{d^{D-1}k}{(2\pi)^{D-1}}\frac{d\omega}{2\pi}\text{exp}\left[-\pi\alpha^{\prime}\tau_2\big(\vec{k}^2+\omega^2 + 2\omega Q\big)\right]\nonumber\\
&&\hspace*{6cm}\times\prod_{n}\frac{1}{(1-q^n)^D(1-\bar{q}^n)^D}\nonumber\\
Z_s(\tau)&=& iV_D|\eta(\tau)|^{-2D}\int\frac{d^{D-1}k}{(2\pi)^{D-1}}\frac{d\omega}{2\pi}\text{exp}\left[-\pi\alpha^{\prime}\tau_2\big(\vec{k}^2+\omega^2 + 2\omega Q + Q^2\big)\right]
\end{eqnarray}
Now just redifing the variable $\omega\rightarrow\omega+Q$ and integrating over the momenta variables we get,
\begin{equation}
Z_s(\tau)=iV_D\left((4\pi^2\alpha^{\prime}\tau_2)^{-\frac{1}{2}}|\eta(\tau)|^{-2}\right)^D=iV_D(Z_X(\tau))^D
\end{equation}
From the fermionic traces we obtain after projecting onto even world-sheet fermion number we get,
\begin{equation}
Z_{\psi}^+=\frac{1}{2}\left(Z^0_{\ 0}(\tau)^{\frac{D}{2}-1}-Z^0_{\ 1}(\tau)^{\frac{D}{2}-1}-Z^1_{\ 0}(\tau)^{\frac{D}{2}-1}-Z^1_{\ 1}(\tau)^{\frac{D}{2}-1}\right)\ .
\end{equation}
Now we know that $Z^1_{\ 1}(\tau)=0$ so,
\begin{equation}
Z_{\psi}^+=\frac{1}{2}\left(Z^0_{\ 0}(\tau)^{\frac{D}{2}-1}-Z^0_{\ 1}(\tau)^{\frac{D}{2}-1}-Z^1_{\ 0}(\tau)^{\frac{D}{2}-1}\right)\ .
\end{equation}
Thus the full partition function is given by\footnote{Here we consider the case of Type IIB theories although generalization to other cases is simple.},
\begin{equation}
Z_{LD}=\int \frac{d\tau d\bar{\tau}}{4\tau_2}iV_D\left((4\pi^2\alpha^{\prime}\tau_2)^{-\frac{1}{2}}|\eta(\tau)|^{-2}\right)^D|\eta(\tau)|^4Z_{\psi}^+(\tau)Z_{\bar{\psi}}^{+}(\bar{\tau})\nonumber
\end{equation}
\begin{equation}
\Rightarrow Z_{LD}=iV_D\int \frac{d\tau d\bar{\tau}}{16\pi^2\alpha^{\prime}\tau_2^2}\left((4\pi^2\alpha^{\prime}\tau_2)^{-\frac{1}{2}}|\eta(\tau)|^{-2}\right)^{D-2}Z_{\psi}^+(\tau)Z_{\bar{\psi}}^{+}(\bar{\tau})
\end{equation}

\subsection{Modular Invariance}
While testing the modular invariance of the above partition function one immediately notices that for arbitrary values of $D>10$ we do not have modular invariance i.e. even though the measure $d\tau d\bar{\tau}/\tau_2^2$ and $Z_X(\tau)$ are modular invariant by themselves $Z^+_{\psi}Z^+_{\bar{\psi}}$ break this invarince. But one can show explicitly that for $D=10+16n$ with integer values of $n$,\footnote{In \cite{Aharony:2006ra} this result was stated and also this result was argued by K. Narayan and collaborators in an unpublished notes.} $Z^+_{\psi}Z^+_{\bar{\psi}}$ is modular invariant owing to the following transformation laws,
\begin{eqnarray}
Z^+_{\psi}(-1/\tau) = Z^+_{\psi}(\tau) &;& Z^+_{\psi}(\tau + 1) = \exp\left(\frac{2\pi(2n+1)i}{3}\right)Z^+_{\psi}(\tau) \\
Z^+_{\bar{\psi}}(-1/\bar{\tau}) = Z^+_{\bar{\psi}}(\bar{\tau}) &\text{and}& Z^+_{\bar{\psi}}(\bar{\tau} + 1) = \exp\left(-\frac{2\pi(2n+1)i}{3}\right)Z^+_{\bar{\psi}}(\bar{\tau})
\end{eqnarray}
Hence for the rest of this paper we take $D=10+16n$. Thus we have including \eqref{eq:D_Q}, 
\begin{equation}
4\alpha^{\prime}Q^2=16n\ ,\quad\Rightarrow\quad Q=2\sqrt{\frac{n}{\alpha^{\prime}}} \ .\label{eq:largeQ}
\end{equation} 

So we see that due to the modular invarince we have,
\begin{equation}
Z_{LD}=iV_D\int_{\mathbb{F}_0} \frac{d\tau d\bar{\tau}}{16\pi^2\alpha^{\prime}\tau_2^2}\left(Z_X(\tau)\right)^{8+16n}Z_{\psi}^+(\tau)Z_{\bar{\psi}}^{+}(\bar{\tau})
\end{equation}
where, $\mathbb{F}_0$ denotes the fundamental domain in the complex $\tau$ upper half plane. Notice that only for $n=0$ i.e. the critical case $Z_{LD}$ vanishes by the use of Riemann Identity. For $n\neq 0$ the partition function is non vanishing which implies that linear dilaton backgrounds break spacetime supersymmetry.

\subsection{Mass renormalization in string pertubation theory}
We know that one has to start with the BV master action when attempting an off-shell formulation of string theory \cite{Zwiebach:1992ie},\cite{Hata:1993gf},\cite{Okawa:2004ii},\cite{Sen:2016bvm},\cite{Berkovits:1995ab},\cite{Witten:1986qs},\cite{Sen:2015hha}. It is then clear that if one is able to solve the resulting equation of motion pertubatively, then there exists a well defined procedure to compute several physical quantities such as mass renormalization \cite{Pius:2013sca},\cite{Pius:2014iaa}, vacuum shift \cite{Pius:2014gza} etc. order by order in perturbation theory. From \cite{Pius:2014iaa} we learn that using the 1PI effective action to some given order in pertubation theory, we have to sum over all the 1PR diagrams for the two point function to get the renormalised mass at that order in the pertubation expansion. The full propagator is given, in accordance with the above statement by,
\begin{equation}
\Pi=\Delta+\Delta\hat{\mathcal{F}}\Delta+\Delta\hat{\mathcal{F}}\Delta\hat{\mathcal{F}}\Delta+....=\Delta(1-\hat{\mathcal{F}}\Delta)^{-1}=(1-\Delta\hat{\mathcal{F}})^{-1}\Delta\ .
\end{equation}
Here $\hat{\mathcal{F}}$ denotes the contribution from the Riemann surfaces that are 1PI in the external momenta. And,
$$\Delta=\frac{1}{2(L_0+\bar{L}_0)}\delta_{L_0,\bar{L}_0}$$
gives the tree level propagator. If $\mathcal{F}$ is the full off-shell two point funtion then,
\begin{equation}
\Pi=\Delta+\Delta\mathcal{F}\Delta.
\end{equation}

Now suppose that we project on to a given mass level $m$, if the full off-shell two point function restricted to this level be denoted by $F_T$ then we have,
\begin{equation}
\tilde{F}_T=F_T(1+(k^2+m^2)^{-1}\mathcal{I}F_T)^{-1}
\end{equation}
with,
$$\mathcal{I}=
\begin{pmatrix}
 & I & \\
 I& &  \\
 & & I 
\end{pmatrix}\ .$$
The propagator restricted to the modified physical sector is given by,
$$\mathcal{V}=U(k)(k^2+m^2-\tilde{F}_d(k))^{-1}U(k)^{\dagger},\ \text{with}\ \tilde{F}(k)=U(k)\tilde{F}_d(k)U(k)^{\dagger}\ \text{and}\ \tilde{F}_{\alpha\beta}(k)=\langle\bar{\alpha}|\tilde{F}_T|\bar{\beta}\rangle_p.$$
So we see that $\tilde{F}(k)$ is the renormalized mass matrix for the physical sector. On diagonalizing we will get the renormalized masses for each physical state. As an exercise we work out the mass renormalization of the massless states like the graviton, in critical superstring theory i.e. $Q=0$ case in Appendix \ref{app:B}. For the critical case, the two point graviton amplitude should vanish since we expect general co-ordinate invarince in the quantum theory of superstrings as well. So the massless states remain massless under quantum corrections. 

\section{Two point amplitude at one loop with $Q\neq 0$}
\label{sec:Q1}
In case of $Q\neq 0$ i.e. the linear dilaton background with world sheet supersymmetry, instead of \eqref{eq:emLD} we get the EM tensor and the world-sheet supercurrent given by,
\begin{eqnarray}
T_B(z)&=&-\frac{1}{\alpha^{\prime}}:\partial X^{\mu}\partial X_{\mu}:\ +\ V_{\mu}\partial^2X^{\mu}\ -\ \frac{1}{2}\psi^{\mu}\partial\psi_{\mu} \\
T_F(z)&=& i(2/\alpha^{\prime})^{1/2}\psi^{\mu}\partial X_{\mu}\ -\ i(2\alpha^{\prime})^{1/2}V_{\mu}\partial\psi^{\mu}
\end{eqnarray}
With these results we can straight forwardly compute the BRST current,
\begin{equation}
Q_B=\frac{1}{2\pi i}\oint (dz j_B(z)\ -\ d\bar{z} \bar{j}_B(\bar{z}) )
\end{equation}
where,
\begin{equation}
j_B=cT_B+\gamma T_F+bc\partial c+\frac{3}{4}\partial c\beta\gamma+\frac{1}{4}c\partial(\beta\gamma)-c\beta\partial\gamma-b\gamma^2 \label{eq:BRST}
\end{equation}
and the same expression with antiholomorphic fields for the antiholomorphic part. 

Let us first find the equation of motion for the first excited level with this BRST current. The on shell condition for an operator $\Phi$, as always is given by, $Q_B\Phi=0$. Since the tachyon has been projected out the first excited level gives the lowest energy i.e. the ground state. The vertex operator for the NS-NS state in the -1 picture number is given by,
\begin{equation}
\bar{c}cV_{-1}(k,z,\bar{z})=\bar{c}c\ e^{-(\phi+\bar{\phi})}\zeta_{\mu\nu}\psi^{\mu}\bar{\psi}^{\nu}e^{ik.X}(z,\bar{z})\label{eq:pic-1}
\end{equation}
Until the next subsection, we concentrate on the holomorphic part only since the anti-holomorphic part works out in a similar fashion. So applying the BRST charge  we get,
\begin{eqnarray}
Q_B (c\ e^{-\phi}\zeta_{\mu\nu}\psi^{\mu}e^{ik.X}(0,0)) &=& \frac{1}{2\pi i}\oint dz\ j_B(z)c\ e^{-\phi}\zeta_{\mu\nu}\psi^{\mu}e^{ik.X}(0,0) \\
&=&\frac{1}{2\pi i}\oint dz\ \frac{(\partial c)c(0)e^{-\phi}\zeta_{\mu\nu}\psi^{\mu}e^{ik.X}(0,0)}{z}\left(\frac{\alpha^{\prime}k^2}{4}+\frac{iV.k\alpha^{\prime}}{2}+\frac{1}{2}\right) \nonumber\\
&+& \frac{1}{2\pi i}\oint dz\ \left(\frac{\alpha^{\prime}}{2}\right)^{1/2}\frac{\eta(0)c(0)e^{ik.X}(0,0)}{z}\zeta_{\mu\nu}(k^{\mu}+2iV^{\mu})\nonumber\\
&+&\frac{1}{2\pi i}\oint dz\ \left(-\frac{1}{2}\right)\frac{(\partial c)c(0)e^{-\phi}\zeta_{\mu\nu}\psi^{\mu}e^{ik.X}(0,0)}{z}\ +\ ....\ .
\end{eqnarray}
The $....$ denote terms which are regular in $z$ and hence they vanish under the closed contour integral. Just to be precise, the first two terms are the contributions due to $cT_B$ and $\gamma T_F$ while the third term is the single pole contribution due to the ghost part. Adding up all the contribution we get the on-shell conditions,
\begin{equation}
k^2+2iV.k=0\ ,\quad\text{and}\quad \zeta_{\mu\nu}(k^{\mu}+2iV^{\mu})=0\ .\label{eq:onshell}
\end{equation}
Notice that if we take $V_{\mu}\rightarrow 0$ we get back the expected on-shell conditions for the massless states in critical superstring theory. For time like $V$ we have, 
\begin{equation}
k^2-2iQk^0=0\ \Rightarrow\ k^2=2iQk^0\ ,\quad\text{and}\quad \zeta_{\mu\nu}k^{\mu}=2iQ\zeta_{0\nu}\ .\label{eq:onshell_tl}
\end{equation}
From the above equation \eqref{eq:onshell_tl} we see that,
\begin{equation}
k^2-2iQk^0=0\ \Rightarrow\ -(k^0+iQ)^2+|\vec{k}|^2=Q^2\ \Rightarrow\ -M^2=Q^2 \label{eq:treeM}
\end{equation}
i.e. the tree level mass squared go as $-Q^2$.

In what follows we take $Q$ to be large. We will work with the convention of $\alpha^{\prime}=2$. Thus \eqref{eq:largeQ} becomes
\begin{equation}
Q=\sqrt{2n}\quad\Rightarrow\ \text{$n\rightarrow\infty$ as $Q\rightarrow\infty$.} \label{eq:largeQ2}
\end{equation}
While computing the two point amplitude on the torus we will focus on the terms that have the highest power of $n$ or equivalently $Q$. 
We will first determine the 0-picture vertex operators and determine first the real part and then the imaginary part of the two-point amplitude.

From the -1 picture vertex in \eqref{eq:pic-1} and the PCO,
\begin{equation}
\chi(z)=e^{\phi}T_F(z)+c\partial\xi(z)+\frac{1}{4}\partial b\eta e^{2\phi}(z)+\frac{1}{4}b(2\partial\eta e^{2\phi}+\eta\partial(e^{2\phi}))(z)\ ,
\end{equation}
we get for the holomorphic part,
\begin{eqnarray}
\chi(z)cV_{-1}(k,0,0)&=& zT_F(z)c\zeta_{\mu}\psi^{\mu}e^{ik.X}(0,0)\ +\ z\partial\xi(z)e^{-\phi}(0)c\zeta_{\mu}\psi^{\mu}e^{ik.X}(0,0)\nonumber\\
&-&\frac{1}{4}z^2.\frac{1}{z^2}\eta(z)e^{\phi}(0)c\zeta_{\mu}\psi^{\mu}e^{ik.X}(0,0)\ +\ \frac{1}{4}2z.\frac{1}{z}\eta(z)e^{\phi}(0)c\zeta_{\mu}\psi^{\mu}e^{ik.X}(0,0)\nonumber\\
&+&\frac{1}{2}z^2.\frac{1}{z}\partial\eta(z)e^{\phi}(0)c\zeta_{\mu}\psi^{\mu}e^{ik.X}(0,0)\ .
\end{eqnarray}
For the coefficient of $z$ we need the single pole contribution for a finite answer in the $z\rightarrow0$ limit. While for the $z^0$ coefficient we need the constant regular piece. Thus in the $z\rightarrow0$ limit we have,
$$V_0(k,0,0)=\zeta_{\mu}\left(i\partial X^{\mu}-\psi^{\mu}(k.\psi)\right)e^{ik.X}(0,0)+\frac{1}{4}\eta e^{\phi}(0)\zeta_{\mu}\psi^{\mu}e^{ik.X}(0,0)\ .$$
The last term above does not conserve the picture number inside the two point one loop amplitude and hence we can drop it. We see that we reach at the same result as in the $Q=0$ case, but now with on shell conditons \eqref{eq:onshell_tl}. And similarly for the anti-holomorphic part.

Now we can compute the two-point amplitide. As in the case with $Q=0$, we have the full amplitude given by,
\begin{equation}
\mathcal{M}_g=g_s^2\int_{\mathbb{F}_0}\frac{d\tau d\bar{\tau}}{4\tau_2}\int d^2z\ \langle b\bar{b}\bar{c}c(0)V_0(k_1,0,0)V_0(k_2,z,\bar{z})\rangle\ .
\end{equation}

\subsection{The real part}
Let us first turn our attention to the real part of the correlator.
\begin{eqnarray}
\left[\int d^2z\ \zeta^{(1)}_{\mu\nu}\zeta^{(2)}_{\rho\sigma}\left\langle\left\lbrace\partial X^{\mu}(0)\bar{\partial}X^{\nu}(0)-\psi^{\mu}(0)(k_1.\psi(0))\bar{\psi}^{\nu}(0)(k_1.\bar{\psi}(0))\right\rbrace\right.\right. \nonumber\\
\left.\left.\times\lbrace\partial X^{\rho}(z)\bar{\partial}X^{\sigma}(\bar{z})-\psi^{\rho}(z)(k_2.\psi(z))\bar{\psi}^{\sigma}(\bar{z})(k_2.\bar{\psi}(\bar{z}))\rbrace e^{ik_1.X}(0,0)e^{ik_2.X}(z,\bar{z})\right\rangle\right. \nonumber\\
\left. -\int d^2z\ \zeta^{(1)}_{\mu\nu}\zeta^{(2)}_{\rho\sigma}\left\langle\left\lbrace \psi^{\mu}(0)(k_1.\psi(0))\bar{\partial}X^{\nu}(0)+\partial X^{\mu}(0)\bar{\psi}^{\nu}(0)(k_1.\bar{\psi}(0))\right\rbrace\right.\right. \nonumber\\
\left.\left.\times\lbrace \psi^{\rho}(z)(k_2.\psi(z))\bar{\partial}X^{\sigma}(\bar{z})+\partial X^{\rho}(z)\bar{\psi}^{\sigma}(\bar{z})(k_2.\bar{\psi}(\bar{z}))\rbrace e^{ik_1.X}(0,0)e^{ik_2.X}(z,\bar{z})\right\rangle\right] \ . \label{eq:RELD}
\end{eqnarray}

We will look at the different terms that contribute to this correlator and focus our attention to the terms that have the highest power of $n$.

We will be using the result,
\begin{align}
\langle e^{ik_1.X}(0,0)e^{ik_2.X}(z,\bar{z})\rangle &=& iV_D(Z_X(\tau))^D(2\pi)^D\delta^D(k_1+k_2)\left|\vartheta_1(z,\tau)\exp\left(-\frac{\pi(\text{Im}\ z)^2}{\tau_2}\right)\right|^{\alpha^{\prime}k_1.k_2}\nonumber\\
&&\times\prod_{i=1}^{2}\left|\frac{2\pi}{\vartheta_1^{\prime}(0,\tau)}\right|^{-\frac{\alpha^{\prime}k_i^2}{2}}\ .
\end{align}
\textbf{Note}: For linear dilaton backgrounds with charge $Q$, the correlation function of tachyon vertices yields, $\delta^D(k_1+k_2+2Q(1-h))$ with $h$ being the genus of the Riemann surface we consider \cite{Ishibashi:2016bno}. In our case we have the torus with $h=1$. Hence the factor of $\delta^D(k_1+k_2)$, which forces $k_2=-k_1$.

Thus applying \eqref{eq:onshell_tl}, \eqref{eq:largeQ2} and $\alpha^{\prime}=2$ we have,
\begin{eqnarray}
\langle e^{ik_1.X}(0,0)e^{ik_2.X}(z,\bar{z})\rangle &=& i(2\pi)^D\delta^D(k_1+k_2)\mathcal{C}_X^{T^2}\left|\frac{2\pi\vartheta_1(z,\tau)}{\vartheta^{\prime}_1(0,\tau)}\exp\left(-\frac{\pi(\text{Im}\ z)^2}{\tau_2}\right)\right|^{-4iQk_1^0}. \nonumber\\ 
&=& i(2\pi)^D\delta^D(k_1+k_2)\mathcal{C}_X^{T^2}\left|\frac{2\pi\vartheta_1(z,\tau)}{\vartheta^{\prime}_1(0,\tau)}\exp\left(-\frac{\pi(\text{Im}\ z)^2}{\tau_2}\right)\right|^{8n} \label{eq:Tachvert}
\end{eqnarray}
With $\mathcal{C}_X^{T^2}=V_D(Z_X(\tau))^D$. Since we are computing everything in a Lorentz's invariant manner we can go to the rest frame of the pseudo-tachyonic mode for which,
$$
(k_1^0)^2=M^2=-Q^2\quad\Rightarrow\quad k_1^0=iQ=i\sqrt{2n}
$$ 
In the rest of this section we will directly state the results for the different correlators we need. The details on how to obtain these results are provided in the Appendix \ref{app:A}.
\begin{itemize}
\item{First let us focus on the piece,
\begin{equation}
\left\langle\partial X^{\mu}(0)\partial X^{\rho}(z)\bar{\partial}X^{\nu}(0)\bar{\partial}X^{\sigma}(\bar{z}) e^{ik_1.X}(0,0)e^{ik_2.X}(z,\bar{z})\right\rangle .
\end{equation}
\begin{align}
=\left[\delta^{\mu\rho}\delta^{\nu\sigma}\left(\frac{(\vartheta_{11}\partial^2_{z}\vartheta_{11}-\partial_z\vartheta_{11}\partial_z\vartheta_{11})(z)}{\vartheta_{11}(z)^2}+\frac{1}{4\pi\tau_2}\right)\left(\frac{(\vartheta_{11}\bar{\partial}^2_{z}\vartheta_{11}-\bar{\partial}_z\vartheta_{11}\bar{\partial}_z\vartheta_{11})(\bar{z})}{\vartheta_{11}(\bar{z})^2}+\frac{1}{4\pi\tau_2}\right)\right. \nonumber\\
\left. +\ \frac{\pi^2}{\tau_2^2}\delta^{\mu\sigma}\delta^{\rho\nu}+\big(\delta^{\mu\rho}k_1^{\sigma}k_2^{\nu}(..)+k_1^{\sigma}k_2^{\mu}\delta^{\rho\nu}(..)\big)+k_1^{\sigma}k_1^{\rho}k_2^{\mu}k_2^{\nu}(..)+....\right]\left\langle\prod_i e^{ik_i.X}\right\rangle.\label{eq:boson}
\end{align}
When we now contract the indeces with $\zeta^{(1)}_{\mu\nu}\zeta^{(2)}_{\rho\sigma}$ and use \eqref{eq:onshell} along with $k_1=-k_2$, we see that the first two terms are $\sim O(1)$, the third term which is quadratic in the momenta contribute at $\sim O(2n)$ and the term which is quartic in momenta contributes at $\sim O(4n^2)$. So we focus on the last piece and it is given by,
\begin{align}
64n^2\zeta^{(1)}_{00}\zeta^{(2)}_{00}\left[\left|\left(\frac{\vartheta^{\prime}_{11}(z)}{\vartheta_{11}(z)}\right)^2-\frac{4\pi^2}{\tau_2^2}(\text{Im}(-z))^2\right|^2+\frac{16\pi^2}{\tau_2^2}\left|\frac{\vartheta^{\prime}_{11}(z)}{\vartheta_{11}(z)}\text{Im}(-z)\right|^2\right.\nonumber\\
\left.+\frac{8\pi}{\tau_2}\text{Im}\left\lbrace\frac{\vartheta^{\prime}_{11}(z)(\bar{\vartheta}^{\prime}_{11}(\bar{z}))^2}{\vartheta_{11}(z)(\bar{\vartheta}_{11}(\bar{z}))^2}\text{Im}(-z)\right\rbrace\right]\left\langle\prod_{i=1,2} e^{ik_i.X}\right\rangle.
\end{align}
This contribution is positive as we can see from the above expression.}
\item{Next we consider the cross terms of the type,
\begin{eqnarray}
\zeta^{(1)}_{\mu\nu}\zeta^{(2)}_{\rho\sigma}\left\langle\partial X^{\mu}(0)\bar{\partial}X^{\nu}(0)\psi^{\rho}(z)(k_2.\psi(z))\bar{\psi}^{\sigma}(\bar{z})(k_2.\bar{\psi}(\bar{z})) e^{ik_1.X}(0,0)e^{ik_2.X}(z,\bar{z})\right\rangle \nonumber\\
=-\zeta^{(1)}_{\mu\nu}\zeta^{(2)}_{\rho\sigma}k_2^{\mu}k_2^{\nu}k_2^{\rho}k_2^{\sigma}\left(\left|\frac{\vartheta_{11}^{\prime}(z)}{\vartheta_{11}(z)}\right|^2+\frac{4\pi^2}{\tau_2^2}(\text{Im}(-z))^2+\frac{4\pi}{\tau_2}\text{Im}\left\lbrace\frac{\vartheta_{11}^{\prime}(z)}{\vartheta_{11}(z)}\text{Im}(-z)\right\rbrace\right)\ \nonumber\\
=\ -64n^2\zeta^{(1)}_{00}\zeta^{(2)}_{00}\left(\left|\frac{\vartheta_{11}^{\prime}(z)}{\vartheta_{11}(z)}\right|^2+\frac{4\pi^2}{\tau_2^2}(\text{Im}(-z))^2+\frac{4\pi}{\tau_2}\text{Im}\left\lbrace\frac{\vartheta_{11}^{\prime}(z)}{\vartheta_{11}(z)}\text{Im}(-z)\right\rbrace\right)\ .
\end{eqnarray}
Hence we see that these terms go as $\sim\ O(Q^4)\ \sim\ O(4n^2)$. And all the cross terms in \eqref{eq:RELD}} come with a minus sign hence these terms also give positive contributions to the amplitude.
\item{The term with four holomorphic and four anti-holomorphic fermions is given by,\\ 
\begin{eqnarray}
k_{1\alpha}k_{2\beta}k_{1\gamma}k_{2\delta}\langle\psi^{\mu}\psi^{\alpha}(0)\psi^{\rho}\psi^{\beta}(z)\bar{\psi}^{\nu}\bar{\psi}^{\gamma}(0)\bar{\psi}^{\sigma}\bar{\psi}^{\delta}(z)e^{ik_1.X}(0,0)e^{ik_2.X}(z,\bar{z})\rangle
\end{eqnarray}
Now for the fermionic correlator we have,
\begin{equation}
k_{1\alpha}k_{2\beta}\langle\psi^{\mu}\psi^{\alpha}(0)\psi^{\rho}\psi^{\beta}(z)\rangle\ =\ k_{1\alpha}k_{2\beta}\left(A_z\delta^{\mu\alpha}\delta^{\rho\beta}+B_z\delta^{\mu\rho}\delta^{\alpha\beta}+C_z\delta^{\mu\beta}\delta^{\alpha\rho}\right)\ . \label{eq:Fcorreln}
\end{equation}
The coefficients can be computed easily and are given by \cite{Sen:2013oza}, \cite{Atick:1986rs},
\begin{eqnarray}
A_z &=& \langle\psi^{1}(0)\psi^{1}(0)\psi^{2}(z)\psi^{2}(z)\rangle\ =\ 1\ ,\\
B_z &=& \langle\psi^{1}(0)\psi^{2}(0)\psi^{1}(z)\psi^{2}(z)\rangle \nonumber\\
&=& -\frac{1}{8(\eta(\tau)^4)^{2n+1}}\left(\frac{\vartheta_{11}^{\prime}(0)}{\vartheta_{11}(z)}\right)^2\sum_{\nu}\delta_{\nu}\left(\vartheta_{\nu}(z)^2+2\vartheta_{\nu}(z)\vartheta_{\nu}(-z)+\vartheta_{\nu}(-z)^2\right)\big(\vartheta_{\nu}(0)^2\big)^{4n+1}\ ,\nonumber\\
\\
C_z &=& \langle\psi^{1}(0)\psi^{2}(0)\psi^{2}(z)\psi^{1}(z)\rangle\ =\ -B_z\ .
\end{eqnarray}
So the full correlator on using \eqref{eq:largeQ} behaves like,
\begin{eqnarray}
\zeta^{(1)}_{\mu\nu}\zeta^{(2)}_{\rho\sigma}\big(A_zk_1^{\mu}k_{2}^{\rho}-B_z(k_1)^2\delta^{\mu\rho}+C_zk_1^{\rho}k_2^{\mu}\big)\big(A_{\bar{z}}k_1^{\nu}k_{2}^{\sigma}-B_{\bar{z}}(k_1)^2\delta^{\nu\sigma}+C_{\bar{z}}k_1^{\sigma}k_2^{\nu}\big)\nonumber\\
= 16n^2\left(4\zeta^{(1)}_{00}\zeta^{(2)}_{00}|A_z+C_z|^2+\zeta^{(1)}_{\mu\nu}\zeta^{(2)\mu\nu}|B_z|^2+4Re[B_z(A_z+C_z)]\zeta^{(1)}_{0\mu}\zeta^{(2)\mu}_{\ 0}\right)\\
\end{eqnarray}
This contribution as we can see gives positive contribution as well and appears in \eqref{eq:RELD} with a $`+$' sign.}
\item{Finally from the last two lines of \eqref{eq:RELD} also, we have negative contributions but since they appear with a $`-$' sign in the equation. So their full contribution to the amplitude is also positive when we keep the highest order terms $n$ i.e. prortional $4n^2$ like in the above cases. The explicit details are given in Appendix \ref{app:A}.}
\end{itemize}

\subsection{The imaginary part}
Let us now look at the contribution from the imaginary part which is given by,
\begin{equation}
\left[\int d^2z\ \zeta^{(1)}_{\mu\nu}\zeta^{(2)}_{\rho\sigma}\left\langle \left\lbrace \psi^{\mu}(0)(k_1.\psi(0))\bar{\partial}X^{\nu}(0)+\partial X^{\mu}(0)\bar{\psi}^{\nu}(0)(k_1.\bar{\psi}(0))\right\rbrace\right.\right. \label{eq:IMLD}
\end{equation}
$$\left.\left.\times\left\lbrace\partial X^{\rho}(z)\bar{\partial}X^{\sigma}(\bar{z})-\psi^{\rho}(z)(k_2.\psi(z))\bar{\psi}^{\sigma}(\bar{z})(k_2.\bar{\psi}(\bar{z}))\right\rbrace e^{ik_1.X}(0,0)e^{ik_2.X}(z,\bar{z})\right\rangle\right.$$
$$\left. +\int d^2z\ \zeta^{(1)}_{\mu\nu}\zeta^{(2)}_{\rho\sigma}\langle (\mu\leftrightarrow\rho,\ \nu\leftrightarrow\sigma,\ k_1\leftrightarrow k_2,\ z,\bar{z}\leftrightarrow 0,0)e^{ik_1.X}(0,0)e^{ik_2.X}(z,\bar{z})\rangle\right]\ .$$

From the first two lines of \eqref{eq:IMLD} we have four terms. We first look at,
\begin{eqnarray}
&& \langle\bar{\partial}X^{\nu}(0)\partial X^{\rho}(z)\bar{\partial}X^{\sigma}(\bar{z})\psi^{\mu}(0)(k_1.\psi(0))e^{ik_1.X}(0,0)e^{ik_2.X}(z,\bar{z})\rangle \\
&& +\ \langle\partial X^{\mu}(0)\partial X^{\rho}(z)\bar{\partial}X^{\sigma}(\bar{z})\bar{\psi}^{\mu}(0)(k_1.\bar{\psi}(0))e^{ik_1.X}(0,0)e^{ik_2.X}(z,\bar{z})\rangle . \nonumber
\end{eqnarray}
Now using \eqref{eq:Bcorreln}, \eqref{eq:Fcorreln} and the onshell condition \eqref{eq:onshell_tl} one can easily check the following facts
\begin{itemize}
\item{The first term has two contributions, one at $\sim O(2n)$ which we will neglect with respect to the contribution at $\sim O(4n^2)$,
$$-64n^2\zeta^{(1)}_{00}\zeta^{(2)}_{00}\left(-i\frac{\vartheta^{\prime}_{11}(z)}{\vartheta_{11}(z)}+\frac{2\pi}{\tau_2}\text{Im}(z)\right)\left(-i\frac{\bar{\vartheta}^{\prime}_{11}(\bar{z})}{\bar{\vartheta}_{11}(\bar{z})}-\frac{2\pi}{\tau_2}\text{Im}(z)\right)\left(i\frac{\bar{\vartheta}^{\prime}_{11}(\bar{z})}{\bar{\vartheta}_{11}(\bar{z})}+\frac{2\pi}{\tau_2}\text{Im}(z)\right)$$
\begin{equation}
\Rightarrow 64n^2\zeta^{(1)}_{00}\zeta^{(2)}_{00}\left(\left|\frac{\vartheta^{\prime}_{11}(z)}{\vartheta_{11}(z)}\right|^2+\frac{4\pi^2}{\tau_2^2}(\text{Im}(z))^2+\frac{4\pi}{\tau_2}\text{Im}\left\lbrace\frac{\vartheta^{\prime}_{11}(z)}{\vartheta_{11}(z)}\text{Im}(z)\right\rbrace\right)\left(i\frac{\bar{\vartheta}^{\prime}_{11}(\bar{z})}{\bar{\vartheta}_{11}(\bar{z})}+\frac{2\pi}{\tau_2}\text{Im}(z)\right)
\end{equation}
Same argument are true for the second term and the contribution at $\sim Q^4$ or $4n^2$ is given by,
$$-64n^2\zeta^{(1)}_{00}\zeta^{(2)}_{00}\left(-i\frac{\vartheta^{\prime}_{11}(z)}{\vartheta_{11}(z)}+\frac{2\pi}{\tau_2}\text{Im}(z)\right)\left(-i\frac{\bar{\vartheta}^{\prime}_{11}(\bar{z})}{\bar{\vartheta}_{11}(\bar{z})}-\frac{2\pi}{\tau_2}\text{Im}(z)\right)\left(i\frac{\vartheta^{\prime}_{11}(z)}{\vartheta_{11}(z)}-\frac{2\pi}{\tau_2}\text{Im}(z)\right).$$
Thus adding them up we get,
\begin{equation}
i64n^2\zeta^{(1)}_{00}\zeta^{(2)}_{00}\left(\left|\frac{\vartheta^{\prime}_{11}(z)}{\vartheta_{11}(z)}\right|^2+\frac{4\pi^2}{\tau_2^2}(\text{Im}(z))^2+\frac{4\pi}{\tau_2}\text{Im}\left\lbrace\frac{\vartheta^{\prime}_{11}(z)}{\vartheta_{11}(z)}\text{Im}(z)\right\rbrace\right)\left(\frac{\bar{\vartheta}^{\prime}_{11}(\bar{z})}{\bar{\vartheta}_{11}(\bar{z})}+\frac{\vartheta^{\prime}_{11}(z)}{\vartheta_{11}(z)}\right). \label{eq:x3psi}
\end{equation}}
\item{And one can check that from the last line in \eqref{eq:IMLD} these contribution which is conjugate to the one described above comes out to be 
$$-i64n^2\zeta^{(1)}_{00}\zeta^{(2)}_{00}\left(\left|\frac{\vartheta^{\prime}_{11}(z)}{\vartheta_{11}(z)}\right|^2+\frac{4\pi^2}{\tau_2^2}(\text{Im}(z))^2+\frac{4\pi}{\tau_2}\text{Im}\left\lbrace\frac{\vartheta^{\prime}_{11}(z)}{\vartheta_{11}(z)}\text{Im}(z)\right\rbrace\right)\left(\frac{\bar{\vartheta}^{\prime}_{11}(\bar{z})}{\bar{\vartheta}_{11}(\bar{z})}+\frac{\vartheta^{\prime}_{11}(z)}{\vartheta_{11}(z)}\right).$$
So when we add this piece to \eqref{eq:x3psi} the total contribution vanishes.}
\item{One can check that similar arguments hold for the other type of term as well i.e. terms with one bosonic field and rest all fermionic fields. When we take the contribution from all possible terms of this type the total contribution vanishes.}
\end{itemize}

One can choose a particular co-ordinate patch i.e. fix some gauge to carry out the integral over the moduli space, but we will not do it in this note.  There is one point we should mention; for the bosonic piece we see explicitly that there are no other divergences other than $z\rightarrow 0$ i.e. they are free from spurious divergences. But for a function like $B_z$ or $C_z$ since we do not completely know zeros and poles for the spin structure sum it is hard to conclude the absence of spurious poles. This can be an interesting direction to check that the perturbative amplitude is indeed completely well defined.

Now from \cite{Pius:2014iaa} we have to get the renomalized mass one has to solve the following equation iteratively,
\begin{equation}
E^2=m^2-\tilde{F}(E)\ ,
\end{equation}  
to a given order in perturbation theory. Here $m^2$ is the tree level mass squared and $\tilde{F}(E)$ is the quantum correction to the propagator. in the current case,
$$m^2=-Q^2\ ,\quad \text{and}\quad \text{Im}(\tilde{F}(E))=0\ ,\quad \tilde{F}(E)>0\ .$$
from the one loop correction. Also since the imaginary part is zero at the order we considered, we have zero decay width so the modes are stable. Hence atleast to this order we can confirm the following statement unambiguously.
\begin{itemize}
\item{The \textit{Pseudotachyon} modes remain pseudotachyonic under quantum corrections and the system stabilizes towards condensation of such modes.}
\end{itemize}

\section{Interpolating functions due to Sen}
\label{sec:IF}
We now attempt to construct an interpolating function which gives the renormalized mass for all coupling and matches the pertubative expansion to a given order at both the weak and strong coupling end if such expansions indeed exists. We will follow the strategy described in the seminal work \cite{Sen:2013oza} for this purpose. We will first briefly discuss the know example where this technique succeeds to within $10\%$ accuracy, and then try to implement it for the Linear Dilaton backgrounds.

\subsection{Example of the heterotic/type I duality}
In case of the Heterotic and type I, S-dual theories the couplings at the two ends are related by,
$$g_Hg_I=2^5\pi^7\ .$$
Thus we can parametrize them as follows,
$$g_H=(2\pi)^{7/2}g,\quad g_I=2^{3/2}\pi^{7/2}g^{-1},$$
so that the two formulae meet at $g=1$. The renormalized mass function of the second excited state\footnote{In our convention we always state the tachyons to be the ground state or zero excitation state, even though in superstring theories they are projected out by GSO projection. } (i.e. the first massive level) was considered. We define, the renormalized mass function $F(g)$ by,
$$M(g)=2^{15/8}\pi^{7/8}F(g)$$ 
For weak coupling end we have the Heterotic theory whereas for the strong coupling end we have the type I theory with,
\begin{equation}
F^W_2(g)=g^{1/4}\left(1+K_wg^2+O(g^4)\right),\ \ F^S_1(g)=g^{3/4}\left(1+K_sg^{-1}+O(g^{-2})\right);\ K_w\simeq 0.23,\ K_s\simeq 0.351
\end{equation}
where one loop or the subleading corrections at both ends where considered.

We now look at the interpolating function which has to agree with the coupling expansions to order $m$ in the weak coupling and order $n$ in strong coupling end.
\begin{equation}
F_{m,n}(g)=g^{1/4}\left[1+a_1g+...+a_mg^m+b_ng^{m+1}+b_{n-1}g^{m+2}+...+b_1g^{m+n}+g^{m+n+1}\right]^{\frac{1}{2(m+n+1)}}\label{eq:Fg}
\end{equation} 
In the present example it is known that for the Heterotic theory being a closed superstring theory has the expansion in $g^2$ so $a_m=0$ for $m$ odd.
The interpolating function which is within $10\%$ accuracy of the actual function was calculated to be,
$$F_{3,1}=g^{1/4}\left(1+10K_wg^2+10K_sg^4+g^5\right)^{1/10}$$

\subsection{Application to the linear dilaton backgrounds}
In our case we would like to apply the same technique to the first excited state of strings which are the \textit{Pseudotachyonic} modes. It is fairly obvious that now all the coefficients for the interpolating function may depend on the background charge $Q$.

\textbf{Note:} It should be understood that this technique is useful only in the cases where it is possible to determine the physical quantities of the theory at both ends perturbatively, with the use of S-duality. In cases where this is not true, this process cannot yield sensible results and one indeed has to carry out a full non-perturbative computation.\\\\

So we will begin with a theory where the weak coupling ($g_w$) regime is S-dual to another theory at the strong coupling ($g_s$) regime. Hence we will have,
\begin{equation}
g_wg_s=\mathcal{C}_0\ ,\quad\text{where $\mathcal{C}_0$ is a constant.} \label{eq:coupling}
\end{equation}
We do the following parametrization,
$$g_w=c_0g,\ \text{and}\ g_s=(\mathcal{C}_0/c_0)g^{-1}$$
such that the formulae meet at $g=1$ i.e. both the strong and weak couplings become of the same order ($\sim\ O(1)$) while satisfying \eqref{eq:coupling}. This parameter now can be treated as,
$$g\equiv g_{st}=e^{-QX^0}\ \ \text{i.e. the coupling varrying with time.}$$
As in the earlier example we define the renormalized mass function via,
$$M(g)=\text{(const.)}\times F(g)\ .$$
Let us first try to see the leading order coefficient of the interpolating function for the case at hand. From \eqref{eq:treeM} we have,
\begin{equation}
k^2-2iQk^0=0\ \Rightarrow\ M^2=-Q^2
\end{equation}
and the leading order coupling behavior at the strong and weak coupling end are assumed to be,
\begin{equation}
F_W(g)\sim g^{\kappa},\quad F_S(g)\sim g^{\delta}\ .
\end{equation}
Note that in case of the example in previous subsection $\kappa=1/4$ and $\delta=3/4$. Here we don't fix the values since they are theory dependent.

With these details, now we can venture to make an ansatz for the interpolating function that agrees the expansions at both ends upto some given order.
\begin{align}
F_{m,n}(g)&=& -Q^2g^{\kappa}\left[1\ +\ a_1(Q)g\ +\ a_2(Q)g^2\ +\ ...+\ a_m(Q)g^m\ +\ b_n(Q)g^{m+1}\ +\ ...\right. \nonumber\\
&&\left. +\ b_1(Q)g^{m+n}\ +\ g^{m+n+1}\right]^{\frac{\delta-\kappa}{m+n+1}}.
\end{align}
We saw that the weak coupling expansion is in powers of $g^2$ since it is a cosed superstring perturbation theory. So $a_m(Q)=0$ for odd $m$. And the coefficient $a_2(Q)\sim O(Q^2)$ in the large $Q$ limit. Similarly at the strong coupling end if the theory is an open superstring theory then $b_1(Q)$ starts at $\sim O(Q^2)$ otherwise, $b_2(Q)$ starts at $\sim O(Q^2)$ and $b_n(Q)=0$ for odd $n$. The above interpolating function matches the leading order expansion at both ends by construction\footnote{In the large $Q$ limit we saw that the corrections go as $\sim\ Q^4$. So if an overall factor of $-Q^2$ is pulled out we are left with a correction factor of $Q^2$ within the parenthesis.}. 
\begin{eqnarray}
\text{Expanding around $g=0$}&:& F_{m,n}(g)\approx -Q^2g^{\kappa}\left(1+\frac{a_2(Q)(\delta - \kappa)}{m+n+1}g^2+O(g^4)\right)\\
\text{Expanding around $g=\infty$}&:& F_{m,n}(g)\approx -Q^2g^{\delta}\left(1+\frac{b_1(Q)(\delta - \kappa)}{m+n+1}g^{-1}+O(g^{-2})\right)
\end{eqnarray}
For the one loop corrected mass at both ends one can compute the interpolating function $F_{3,1}$ (like in the previous example) we have, 
\begin{equation}
F_{3,1}(g)=-Q^2g^{\kappa}\left(1\ +\ a_2Q^2g^2\ +\ b_1Q^2g^4\ +\ g^5\right)^{(\delta-\kappa)/5}.
\end{equation}
We have written, 
$$a_2(Q)\equiv a_2Q^2,\quad b_1(Q)\equiv b_1Q^2,\ \text{with}\ a_2>0,\ b_1>0$$
$$\text{in accordance with their large $Q$ expansion.}$$\\
Now that we have a function which gives the renormalized mass for the lowest energy state we can take the limit $g\rightarrow 1$ i.e. at time $X^0=0$ to find the lowest energy state to which the universe tunnels to from early time strongly coupled region in case of the linear dilaton (time dependent) background. Hence after tunneling we reach a capping state configuration with the lowest energy state given by,
\begin{equation}
M_{cs}= -Q^2\left[2\ +\ Q^2(a_2\ +\ b_1)\right]^{(\delta-\kappa)/5}\times (\text{const.}).
\end{equation} 
The over all constant is also theory specific like $\kappa$ and $\delta$. The accuracy of the result obviously depends on the accuracy of the interpolating function and it increases or stabilizes as we go to higher order in pertubation theory at both ends.

\section{Discussions}
\label{disc}
In this note we have presented a strategy of constructing the so called ``capping" state in SCLD theories by embedding the SCLD phase in a strong coupling completion. Although we are unable to produce exact numbers by working with some specific theories, we do give consistency arguments to support that the strategy employed may give sensible results. Let us list the results we obtained for summarising this work,
\begin{itemize}
\item{We show for the two point function on the torus, that in the large $Q$ limit the imaginary part vanishes to the highest order in $Q$ implyng a vanishing decay width to this order. From the real part we get a positive contribution so that the energy of the Pseudo-tachyonic mode lowers under the one loop correction of the mass.}
\item{We find the interpolating function which is supposed to match at both the weak and strong coupling end in case of theories which are S-dual to each other.}
\end{itemize}

An interesting future direction is to actually compute the fermionic correlation function using some mathematical tools to figure out whether there are any spurious poles for this amplitude. Another open problem is to find specific theories satisfying the relevant conditions so that this procedure can be checked explicitly. But these are currently at the level of speculation. If we do infact find specific theories then we can produce numbers which will render a concrete example where this strategy can be tested for determining the initial capping state within some reasonable accuracy. 

\section*{Acknowledgement}
I would like to thank Ashoke Sen, Harold Erbin and K. Narayan for their useful comments on the earlier versions of the draft which improved my understanding significantly.
\appendix

\section{Relevant correlators for $Q\neq 0$}
\label{app:A}
\begin{enumerate}
\item{Correlation function due to the bososnic piece is given by ($\alpha^{\prime}=2$)\footnote{Since, $\vartheta_{11}(-z)=-\vartheta_{11}(z)$, hence $\left(i\partial_z\text{ln}[\vartheta_{11}(-z)]+\frac{2\pi}{\tau_2}\text{Im}(-z)\right)=\left(i\partial_z\text{ln}[\vartheta_{11}(z)]-\frac{2\pi}{\tau_2}\text{Im}(z)\right)$.},
$$\left\langle\partial X^{\mu}(0)\partial X^{\rho}(z)\bar{\partial}X^{\nu}(0)\bar{\partial}X^{\sigma}(\bar{z}) e^{ik_1.X}(0,0)e^{ik_2.X}(z,\bar{z})\right\rangle$$
\begin{align}
=\ \delta^{\mu\rho}\left(\frac{(\vartheta_{11}\partial^2_{z}\vartheta_{11}-\partial_z\vartheta_{11}\partial_z\vartheta_{11})(z)}{\vartheta_{11}(z)^2}+\frac{1}{4\pi\tau_2}\right)\langle\bar{\partial}X^{\nu}(0)\bar{\partial}X^{\sigma}(z)e^{ik_1.X}(0,0)e^{ik_2.X}(z,\bar{z})\rangle \nonumber\\
+k_2^{\mu}\left(i\partial_z\text{ln}[\vartheta_{11}(z)]-\frac{2\pi}{\tau_2}\text{Im}(z)\right)\langle\partial X^{\rho}(z)\bar{\partial}X^{\nu}(0)\bar{\partial}X^{\sigma}(z)e^{ik_1.X}(0,0)e^{ik_2.X}(z,\bar{z})\rangle \nonumber \\
+\frac{\pi}{\tau_2}\delta^{\mu\sigma}\langle\partial X^{\rho}(z)\bar{\partial}X^{\nu}(0)e^{ik_1.X}(0,0)e^{ik_2.X}(z,\bar{z})\rangle\ . \label{eq:Bcorreln}
\end{align}
If we further evaluate the correlators for each term it is fairly obvious as to what will be the expansion in powers of $\alpha^{\prime}/2$ and $Q$ once we apply \eqref{eq:onshell_tl},
\begin{align}
\sim\ \left\lbrace O\left(\frac{\alpha^{\prime 2}}{4}\right)+O\left(\frac{\alpha^{\prime 3}Q^2}{8}\right)+O\left(\frac{\alpha^{\prime 4}Q^4}{16}\right)\right\rbrace
\end{align}
}
\item{We now look at the fermionic correlator to determine $B_z$, given by,
\begin{eqnarray}
B_z &=& \langle\psi^{1}(0)\psi^{2}(0)\psi^{1}(z)\psi^{2}(z)\rangle\nonumber\\
&=& -\frac{1}{4}\left(\langle\chi^1(0)\tilde{\chi}^1(z)\chi^2(0)\tilde{\chi}^2(z)\rangle\ +\ \langle\tilde{\chi}^1(0)\chi^1(z)\chi^2(0)\tilde{\chi}^2(z)\rangle\right.\nonumber \\
&&+\ \left.\langle\chi^1(0)\tilde{\chi}^1(z)\tilde{\chi}^2(0)\chi^2(z)\rangle\ +\ \langle\tilde{\chi}^1(0)\chi^1(z)\tilde{\chi}^2(0)\chi^2(z)\rangle\right)\ .
\end{eqnarray}
Although the expression for $\langle\chi(0)\tilde{\chi}(z)\rangle,$ and $ \langle\tilde{\chi}(0)\chi(z)\rangle$ obeys \eqref{eq:cmplxF}, but now taking in contribution from all the fermions and superconformal ghosts we get (with $D=10+16n$),
\begin{equation}
B_z=-\frac{1}{8(\eta(\tau)^4)^{2n+1}}\left(\frac{\vartheta_{11}^{\prime}(0)}{\vartheta_{11}(z)}\right)^2\sum_{\nu}\delta_{\nu}\left(\vartheta_{\nu}(z)^2+2\vartheta_{\nu}(z)\vartheta_{\nu}(-z)+\vartheta_{\nu}(-z)^2\right)\big(\vartheta_{\nu}(0)^2\big)^{4n+1}.
\end{equation}}
\item{Let us now consider the last two lines of \eqref{eq:RELD}. There are two types of terms that we need to consider if we just want to check their expansion in powers of $\alpha^{\prime}$ and $Q$.\\
Firstly we have,
\begin{equation}
k_{1\alpha}k_{2\beta}\langle\psi^{\mu}\psi^{\alpha}(0)\psi^{\rho}\psi^{\beta}(z)\bar{\partial}X^{\nu}(0)\bar{\partial}X^{\sigma}(z)e^{ik_1.X}(0,0)e^{ik_2.X}(z,\bar{z})
\end{equation}
and,
\begin{equation}
k_{1\alpha}k_{2\beta}\langle\partial X^{\mu}(0)\partial X^{\rho}(z)\bar{\psi}^{\nu}\bar{\psi}^{\alpha}(0)\bar{\psi}^{\sigma}\bar{\psi}^{\beta}(z)e^{ik_1.X}(0,0)e^{ik_2.X}(z,\bar{z})\ .
\end{equation}
Using \eqref{eq:Fcorreln}, \eqref{eq:Bcorreln} and of course \eqref{eq:onshell_tl} we get the behavior of these type of terms in the large $Q$ limt to be negative and proportional to $4n^2$. And they occur in the amplitide with a `$-$' sign, so contribution to the amplitude is positive.
\\\\
Lastly we have the terms of type,
\begin{eqnarray}
k_{1\alpha}k_{2\beta}\left\langle\partial X^{\rho}(z)\bar{\partial}X^{\nu}(0)\psi^{\mu}\psi^{\alpha}(0)\bar{\psi}^{\sigma}\bar{\psi}^{\beta}(\bar{z}) e^{ik_1.X}(0,0)e^{ik_2.X}(z,\bar{z})\right\rangle \\
\nonumber\\
\text{and}\quad k_{1\alpha}k_{2\beta}\left\langle\partial X^{\mu}(0)\bar{\partial}X^{\sigma}(\bar{z})\psi^{\rho}\psi^{\beta}(z)\bar{\psi}^{\nu}\bar{\psi}^{\alpha}(0) e^{ik_1.X}(0,0)e^{ik_2.X}(z,\bar{z})\right\rangle
\end{eqnarray}
Using the same set of equations one can again show that the same argument holds.

But now we can see clearly that the piece which proprotional to $2n$ is completely regular in $z$ and $\bar{z}$ while the term proportional to $4n^2$ has a single pole for both $z$ and $\bar{z}$. Hence when we carry out the integration over $z$ plane the first term vanishes while for the second term we can pick up the residue at $z=\bar{z}=0$. So these type of terms are free from spurious poles and are proportional to $4n^2$.}
\end{enumerate}

\section{Two point graviton one loop amplitude with $Q=0$}
\label{app:B}

For this computation we need a -1 picture unintegrated vertex whose location on the torus can be fixed (say (0,0)) by the CKV of the torus and a 0 picture integrated vertex operator and a PCO. The following definitions will be useful in our case,
\begin{eqnarray}
\text{-1 picture vertex :}&& V_{-1}(k,z,\bar{z})=i\ \bar{c}c e^{-(\phi+\bar{\phi})}\zeta_{\mu\nu}\psi^{\mu}(z)\bar{\psi}^{\nu}(\bar{z})e^{ik.X}(z,\bar{z})\\
\text{World-sheet supercurrent :}&& T_F(z)=i(2/\alpha^{\prime})^{\frac{1}{2}}\psi^{\rho}\partial X_{\rho}\ ,\ \bar{T}_F(\bar{z})=i(2/\alpha^{\prime})^{\frac{1}{2}}\bar{\psi}^{\rho}\bar{\partial}X_{\rho}\\
\text{PCO :}&& \chi = \oint dz\ e^{\phi}T_F(z)+...\ ,\quad \bar{\chi} = \oint d\bar{z}\ e^{\bar{\phi}}\bar{T}_F(\bar{z})+...
\end{eqnarray}
The $...$ denotes involving $\eta,\ \partial\xi$ etc. We will not be needing those for our computation.

When $\chi\bar{\chi}$ hits $V_{-1}(k_1,0,0)$ is converted to a 0 picture unintegrated vertex given by,
\begin{equation}
\bar{c}cV_0(k_1,0,0)=\left(\frac{2}{\alpha^{\prime}}\right)\bar{c}c\zeta^{(1)}_{\mu\nu}\left\lbrace\partial X^{\mu}+i\left(\frac{\alpha^{\prime}}{2}\right)\psi^{\mu}(k_1.\psi)\right\rbrace\left\lbrace\bar{\partial} X^{\nu}+i\left(\frac{\alpha^{\prime}}{2}\right)\bar{\psi}^{\nu}(k_1.\bar{\psi})\right\rbrace e^{ik_1.X}(0,0)
\end{equation}
while the 0 picture integrated vertex is given by,
\begin{equation}
U_0(k_2)=\left(\frac{2}{\alpha^{\prime}}\right)\int d^2z\ \zeta^{(2)}_{\mu\nu}\lbrace\partial X^{\mu}+i\psi^{\mu}(k_2.\psi)\rbrace\lbrace\bar{\partial} X^{\nu}+i\bar{\psi}^{\nu}(k_2.\bar{\psi})\rbrace e^{ik_2.X}(z,\bar{z})
\end{equation}
The correlation function that we will be computing is,\footnote{We will be using the convention $\alpha^{\prime}=2$ for the rest of this appendix.}
$$\int_{\mathbb{F}_0}\frac{d\tau d\bar{\tau}}{4\tau_2}\langle b\bar{b}(0)\bar{c}c(0)V_0(k_1,0,0)U_0(k_2)\rangle\ .$$

One can immediately see that there are four possible terms for each of the the real and the imaginary parts. We will look at a few of these terms and guess the result for the rest from the simarities. The correlation function due to the $bc$ ghost zero mode has the well known behavior,
\begin{equation}
\langle b(0)\bar{b}(0)\bar{c}(0)c(0)\rangle_{bc}\ =\ |\eta(\tau)|^4\ .
\end{equation}

Let us now first look at the imaginary part of the matter correlation function, 
\begin{equation}
\int d^2z\ \zeta^{(1)}_{\mu\nu}\zeta^{(2)}_{\rho\sigma}\langle\lbrace \psi^{\mu}(0)(k_1.\psi(0))\bar{\partial}X^{\nu}(0)+\partial X^{\mu}(0)\bar{\psi}^{\nu}(0)(k_1.\bar{\psi}(0))\rbrace \label{eq:IMgr}
\end{equation}
$$\times\lbrace\partial X^{\rho}(z)\bar{\partial}X^{\sigma}(\bar{z})-\psi^{\rho}(z)(k_2.\psi(z))\bar{\psi}^{\sigma}(\bar{z})(k_2.\bar{\psi}(\bar{z}))\rbrace e^{ik_1.X}(0,0)e^{ik_2.X}(z,\bar{z})\rangle$$
$$+\int d^2z\ \zeta^{(1)}_{\mu\nu}\zeta^{(2)}_{\rho\sigma}\langle (\mu\leftrightarrow\rho,\ \nu\leftrightarrow\sigma,\ k_1\leftrightarrow k_2,\ z,\bar{z}\leftrightarrow 0,0)e^{ik_1.X}(0,0)e^{ik_2.X}(z,\bar{z})\rangle\ .$$

Once we compute the first term the second one should work out in a similar manner with the relevant exchanging of indices. The tensor structure for this part is given by,
\begin{equation}
\zeta^{(1)}_{\mu\nu}\zeta^{(2)}_{\rho\sigma}\left(\delta^{\mu\alpha}k_{1\alpha}\langle\bar{\partial}X^{\nu}\mathcal{K}^{\rho\sigma}e^{ik_1.X}(0,0)e^{ik_2.X}(z,\bar{z})\rangle+k_{1\alpha}\delta^{\nu\alpha}\langle\partial X^{\mu}\mathcal{K}^{\rho\sigma}e^{ik_1.X}(0,0)e^{ik_2.X}(z,\bar{z})\rangle\right)\ ,
\end{equation}
where,
$$\mathcal{K}^{\rho\sigma}=\partial X^{\rho}(z)\bar{\partial}X^{\sigma}(\bar{z})-A\ k_{2\alpha}k_{2\beta}\delta^{\rho\alpha}\delta^{\sigma\beta}\ .$$
So we see that both contributions are proportional to,
$$\text{either}\ k_1^{\mu}\zeta^{(1)}_{\mu\nu}\ \text{or}\ k_1^{\nu}\zeta^{(1)}_{\mu\nu}$$
Thus for the graviton on-shell condition these contributions vanish. 
Similarly the second term in \eqref{eq:IMgr} also vanishes which suggests that the imaginary part of the amplitude is zero.

Now we turn our attention to the real part of the amplitude. This part has contributions from the following terms,
$$\int d^2z\ \zeta^{(1)}_{\mu\nu}\zeta^{(2)}_{\rho\sigma}\langle\lbrace\partial X^{\mu}(0)\bar{\partial}X^{\nu}(0)-\psi^{\mu}(0)(k_1.\psi(0))\bar{\psi}^{\nu}(0)(k_1.\bar{\psi}(0))\rbrace$$
$$\times\lbrace\partial X^{\rho}(z)\bar{\partial}X^{\sigma}(\bar{z})-\psi^{\rho}(z)(k_2.\psi(z))\bar{\psi}^{\sigma}(\bar{z})(k_2.\bar{\psi}(\bar{z}))\rbrace e^{ik_1.X}(0,0)e^{ik_2.X}(z,\bar{z})\rangle$$
$$-\int d^2z\ \zeta^{(1)}_{\mu\nu}\zeta^{(2)}_{\rho\sigma}\langle\lbrace \psi^{\mu}(0)(k_1.\psi(0))\bar{\partial}X^{\nu}(0)+\partial X^{\mu}(0)\bar{\psi}^{\nu}(0)(k_1.\bar{\psi}(0))\rbrace$$
\begin{equation}
\times\lbrace \psi^{\rho}(z)(k_2.\psi(z))\bar{\partial}X^{\sigma}(\bar{z})+\partial X^{\rho}(z)\bar{\psi}^{\sigma}(\bar{z})(k_2.\bar{\psi}(\bar{z}))\rbrace e^{ik_1.X}(0,0)e^{ik_2.X}(z,\bar{z})\rangle \ .\label{eq:REgr}
\end{equation}
For the second piece we see that only terms with four fermions all of which are either left or right moving, contribute. Other terms vanish on-shell by previous arguments. So we need to compute the correlation function,
\begin{equation}
\langle\psi^{\mu}(0)\psi^{\alpha}(0)\psi^{\rho}(z)\psi^{\beta}(z)\rangle\equiv\ A\delta^{\mu\alpha}\delta^{\rho\beta}\ +\ B\delta^{\mu\rho}\delta^{\alpha\beta}\ +\ C\delta^{\mu\beta}\delta^{\alpha\rho}
\end{equation}
The first term in the above equation vanishes due to previous arguments. Although the second and third term cannot be determined by these arguments. The correlation function for such tensor structures receive contributions from terms of the form,
\begin{equation}
\langle\psi^i(0)\psi^j(0)\psi^i(z)\psi^j(z)\rangle\ \text{or}\ \langle\psi^i(0)\psi^j(0)\psi^j(z)\psi^i(z)\rangle\quad\text{with}\ \ i\neq j\ .
\end{equation}
Let us look at the cases where $i$, $j$ takes values 1, 2, 3, 4. Here we keep in mind that $\psi^0$ is wick rotataed to $\psi^4$ at the cost of picking up a factor of $i$, and also we combine the fermions into complex pairs via,
$$\chi_k=\frac{1}{\sqrt{2}}\left(\psi^k+i\psi^{k+4}\right),\quad \tilde{\chi}^k=\frac{1}{\sqrt{2}}\left(\psi^k-i\psi^{k+4}\right),\quad 1\leq k\leq 4\ .$$
As a concrete example let us look at the following possible contribution\footnote{One can easily check that for, $i=k$ and $j=k+4$ the total contribution always vanishes in this case.},
\begin{align}
\langle\psi^1(0)\psi^2(0)\psi^1(z)\psi^2(z)\rangle &=& -\frac{1}{4}\left(\langle\chi^1(0)\tilde{\chi}^1(z)\chi^2(0)\tilde{\chi}^2(z)\rangle\ +\ \langle\tilde{\chi}^1(0)\chi^1(z)\chi^2(0)\tilde{\chi}^2(z)\rangle\right.\nonumber \\
&&+\ \left.\langle\chi^1(0)\tilde{\chi}^1(z)\tilde{\chi}^2(0)\chi^2(z)\rangle\ +\ \langle\tilde{\chi}^1(0)\chi^1(z)\tilde{\chi}^2(0)\chi^2(z)\rangle\right)\ .
\end{align}
Using the results of \cite{Atick:1986rs}, \cite{Sen:2013oza} we have for a single complex fermion $\chi^i$,
\begin{equation}
\langle\chi^i(0)\tilde{\chi}^i(z)\rangle\ =\ \frac{1}{\eta(\tau)}\frac{\vartheta^{\prime}_{11}(0)}{\vartheta_{11}(z)}\vartheta_{\nu}(z)\ ,\quad \langle\tilde{\chi}^i(0)\chi^i(z)\rangle\ =\ \frac{1}{\eta(\tau)}\frac{\vartheta^{\prime}_{11}(0)}{\vartheta_{11}(z)}\vartheta_{\nu}(-z)\ .\label{eq:cmplxF}
\end{equation}

If we now take contributions from all the holomorphic fermions and the superconformal ghosts we end up with,
\begin{align}
\langle\psi^1(0)\psi^2(0)\psi^1(z)\psi^2(z)\rangle &=&-\frac{1}{8\eta(\tau)^4}\left(\frac{\vartheta_{11}^{\prime}(0)}{\vartheta_{11}(z)}\right)^2\sum_{\nu}\delta_{\nu}\left(\vartheta_{\nu}(z)^2\vartheta_{\nu}(0)^2+2\vartheta_{\nu}(z)\vartheta_{\nu}(-z)\vartheta_{\nu}(0)^2\right. \nonumber\\
&& \left. +\vartheta_{\nu}(-z)^2\vartheta_{\nu}(0)^2 \right).\nonumber\\ \label{eq:frmn}
\end{align}
The factor of half from the GSO projection is accounted for within the overall factor $1/8$. The r.h.s. of equation \eqref{eq:frmn} vanishes by the use of the Riemann Identity,
\begin{align}
\sum_{\nu}\delta_{\nu}\vartheta_{\nu}(z_1)\vartheta_{\nu}(z_2)\vartheta_{\nu}(z_3)\vartheta_{\nu}(z_4) &=& 2\vartheta_{11}\left((z_1+z_2+z_3+z_4)/2\right)\vartheta_{11}\left((z_1+z_2-z_3-z_4)/2\right) \nonumber\\
&&\vartheta_{11}\left((z_1-z_2-z_3+z_4)/2\right)\vartheta_{11}\left((z_1-z_2+z_3-z_4)/2\right)\ .\nonumber\\
\end{align}

Thus we see that the only possible term that might contribute to \eqref{eq:REgr} is,
\begin{equation}
\langle\partial X^{\mu}(0)\partial X^{\rho}(z)\bar{\partial}X^{\nu}(0)\bar{\partial}X^{\sigma}(z)e^{ik_1.X}(0,0)e^{ik_2.X}(z,\bar{z})\rangle\ .
\end{equation}
Following [ref] we find that this is given by,
\begin{align}
\delta^{\mu\rho}\left(\frac{(\vartheta_{11}\partial^2_{z}\vartheta_{11}-\partial_z\vartheta_{11}\partial_z\vartheta_{11})(z)}{\vartheta_{11}(z)^2}+\frac{1}{4\pi\tau_2}\right)\langle\bar{\partial}X^{\nu}(0)\bar{\partial}X^{\sigma}(z)e^{ik_1.X}(0,0)e^{ik_2.X}(z,\bar{z})\rangle \nonumber\\
+k_2^{\mu}\left(i\partial_z\text{ln}[\vartheta_{11}(-z)]+\frac{2\pi}{\tau_2}\text{Im}(-z)\right)\langle\partial X^{\rho}(z)\bar{\partial}X^{\nu}(0)\bar{\partial}X^{\sigma}(z)e^{ik_1.X}(0,0)e^{ik_2.X}(z,\bar{z})\rangle \nonumber \\
-\frac{\pi}{\tau_2}\delta^{\mu\sigma}\langle\partial X^{\rho}(z)\bar{\partial}X^{\nu}(0)e^{ik_1.X}(0,0)e^{ik_2.X}(z,\bar{z})\rangle\ .
\end{align}
Notice that both the first piece and the second piece have double poles at $z=0$ owing to the fact that the correlation function due to the exponentials give,
$$
\langle e^{ik_1.X}(0,0)e^{ik_2.X}(z,\bar{z})\rangle\sim (2\pi)^{10}\delta^{10}(k_1+k_2)\left|\vartheta_1(z,\tau)\exp\left(-\frac{\pi(\text{Im}\ z)^2}{\tau_2}\right)\right|^{\alpha^{\prime}k_1.k_2}\prod_{i=1}^{2}\left|\frac{2\pi}{\vartheta_1^{\prime}(0,\tau)}\right|^{\frac{-\alpha^{\prime}k_i^2}{2}}
$$
and due to the delta function we get $k_1=-k_2$ which implies $k_1.k_2=-k_1^2=-k_2^2=0$ from the on-shell condition. Thus,
$$\langle e^{ik_1.X}(0,0)e^{ik_2.X}(z,\bar{z})\rangle\sim (2\pi)^{10}\ .$$
We have an integration over $z$ since one vertex is an integrated vertex operator. Near the pole we see the behavior,
\begin{equation}
\oint \frac{dz}{z^2}(2\pi)^{10}f(k,\zeta)\quad\text{where $f(k,\zeta)$ is a function of polarization and momenta.}
\end{equation}
which vanishes from Cauchy's integral formulae. Where as terms proportional to $1/4\pi\tau_2$ or $\pi^2/\tau_2^2$ vanish since they have no singularities on the torus described by the $z$-coordinate. Terms which have a single pole at $z=0$ survive at the end and it is clear from the above expression that they are proportional to,
$$k_2^{\mu}\zeta^{(1)}_{\mu\nu},\ k_1^{\rho}\zeta^{(2)}_{\rho\sigma},\ \text{or}\ k_1^{\rho}\zeta^{(2)}_{\rho\sigma}\zeta^{(1)\mu}_{\ \nu}$$
Now from the correlation function of the exponentials we obtain an overall momentum conserving delta function $\delta^{10}(k_1+k_2)$ so that $k_2=-k_1$, and we get zero contributions imposing the on-shell conditions.

Thus we see that the graviton two point one loop amplitude vanishes on-shell as we had expected.

\end{document}